# Ability and Context Based Adaptive System: A Proposal for Machine Learning Approach

**Elgin Akpınar**
Middle East Technical University
Çankaya/Ankara, Turkey
elgin.akpinar@metu.edu.tr

**Yeliz Yeşilada**
Middle East Technical University Northern
Cyprus Campus
Güzelyurt, Mersin 10, Turkey
yesilady@metu.edu.tr

**Selim Temizer**
Middle East Technical University
Çankaya/Ankara, Turkey
temizer@ceng.metu.edu.tr

## ABSTRACT

When we interact with small screen devices, sometimes we make errors, due to our abilities/disabilities, contextual factors that distract our attention or problems related to the interface. Recovering from these errors may be time consuming or cause frustration. Predicting and learning these errors based on the previous user interaction and contextual factors, and adapting user interface to prevent from these errors can improve user performance and satisfaction. In this paper, we propose a system that aims to monitor user performance and contextual changes and do adaptations based on the user performance by using machine learning techniques. Here, we briefly present our systematic literature review findings and discuss our research questions towards developing such an adaptive system.

## CCS CONCEPTS

• **Human-centered computing** → **HCI design and evaluation methods**; **Empirical studies in HCI**; **Ubiquitous and mobile computing design and evaluation methods**; **Empirical studies in ubiquitous and mobile computing**; *Interaction techniques*; *Ubiquitous and mobile devices.*

## KEYWORDS

Context, small screen devices, wearable devices

**ACM Reference Format:**
Elgin Akpınar, Yeliz Yeşilada, and Selim Temizer. 2019. Ability and Context Based Adaptive System: A Proposal for Machine Learning Approach. In *CHI'19: ACM CHI Conference on Human Factors in Computing Systems, May 04–09, 2019, Glasgow, UK*. ACM, New York, NY, USA, 8 pages. https://doi.org/10.1145/1122445.1122456

## INTRODUCTION

Over the past decade, small screen devices have been a major part of our everyday life. Along with communication, they are used to perform most of our daily tasks [3]. These devices range from tablets, smartphones, phablets[1] to wearable devices such as smart watches and glasses [25]. One of the common characteristics of these devices is their small screen size relative to desktop and laptop computers. Although this feature comes with portability aspect, it significantly affects how we interact with these device, especially from the usability perspective [15, 34].

[1] devices with capabilities of both tablet and smartphones

Desktop computers are typically used in a fixed and stable environment; for instance, a typical setting would be the user seated with no excessive light or weather conditions. Small screen devices, on the other hand, can be used almost anywhere, including indoors, outdoors, noisy, quite or crowded environments. Furthermore, while using small screen devices, the users might be engaged with different and/or parallel tasks, for example messaging while walking on a busy street [23]. In the literature, these types of temporary reductions in user performance due to context are referred to as situationally-induced impairments and disabilities (SIIDs) [33]. This phenomenon was defined as "difficulty accessing computers due to the context or situation one is in, as opposed to a physical impairment" [27, 28]. There can be many factors causing SIIDs and the main observation is that both the environment and the current context can cause SIIDs. In this article, we use context to refer to both environment, situation and context which is defined as "any information that characterizes a situation related to the interaction between humans, applications and the surrounding environment" [8].

These situational impairments can be reduced with adaptive systems, which change themselves with respect to context or user behavior [33]. According to this approach, users do not adapt themselves to a system; instead, system measures performance of the users and adapt itself based on the users' performance. For example, if user has problems with clicking on a target, system may increase the size of the target, or adjust the mouse settings to prevent the error. In this study, we aim to develop an intelligent system using machine learning algorithms which learns and predicts user performance with respect to contextual factors and user interaction; and adapts itself to user and context. In this paper, we briefly present our systematic literature review findings and discuss our research questions to develop an adaptive system.

## SYSTEMATIC REVIEW

In order to identify the research gaps in the field, we have conducted a systematic review of the work that has been done on SIIDs and the impact they showed on small screen and wearable device users' performance. In particular, we asked the following two research questions: (1) "Which contextual factors have been examined in the literature for small screen or wearable device interaction that can cause SIIDs?" and (2) "How do different contextual factors affect small screen or wearable device users' performances?". Based on these research questions, we have reviewed a wide range of articles from online platforms, popular HCI conferences and journals. In order to present the contextual factors, we used the context framework proposed by Jumisko-Pyykkö and Vainio [19] as the backbone of our review.

We have classified the contextual factors used in the literature into physical, temporal, social, task and technical context dimension based on our first research question. Our review has shown that, physical context such as mobility or location has been widely studied. On the other hand, far too little attention has been paid to other contextual factors such as social and temporal. One major problem with the majority of the studies on the field is that, they have been conducted in controllable and maintainable environmental settings. For instance, researchers have preferred laboratories or controlled environments, where participants were isolated from external distractions [4, 20] and environmental factors such as light condition stayed the same between sessions [12]. Moreover, participants have completed experimental tasks with respect to the aim of the studies. Most of these tasks have been completed by using experimental applications, which are simplistic compared to real world applications [4]. The duration of the experiment sessions have been fixed and included time required for participants to learn how to use the application or perform a gesture. As a result, very little is currently known about the effect of contextual factors on users' performance in real world settings [1, 11].

Our investigation on evaluation metrics and contextual factors used to compare user performance has shown that, conditions that can be easily simulated and ensure identical experiments have been used. Two popular conditions are mobility (walking on a treadmill [16], walking on a predefined route [7] or walking freely [31]) and encumbrance [10, 22]. On the other hand, other contextual factors such as lighting level [2], temperature [26], ambient noise [17] or people around interaction [14, 18] are hard to control and they can easily differ between sessions. As a result, only a few studies have used these conditions to compare user performance. Mobility has been a popular contextual factor; however, its effect on user performance have been inconsistent and contradictory. Although many researchers agree on the negative effect of using a mobile device on user posture and attention to surroundings [20, 29]; there is no consensus on the effect of mobility on user performance. For instance, [32] and [9] have found significant effects of mobility on user performance; while [24] could not find a main effect of mobility.

## A PROPOSAL FOR MACHINE LEARNING APPROACH

Unlike environmental settings in the literature, real world use cases include various lighting levels [12], complex obstacles or disruptions [1, 4, 6, 20], environmental noise [30] and any factor that requires attentional resources for safety reasons [5]. Moreover, some of the everyday tasks require continuous attention [23] or are cognitively demanding [21]. Therefore, further experiments should be conducted in users' own environments and with users' own tasks to better understand the effect of contextual factors on user performance.

Overall goal of this study is to develop a system that learns how users interact with small screen devices with respect to contextual factors and predict when users make errors or have usability problems. Based on this prediction mechanism, we plan to build an adaptive system which adapt itself to users and context. In order to reflect real world use cases, we are going to conduct in-situ experiments to collect data from users in their own environments.

### Research Questions

In the rest of this study, we plan to investigate the following research questions:

*How contextual factors affect user performance in terms of error rate and task completion time on small screen devices? Is there a correlation between contextual factors and user performance?* The first part of the study focuses on the effect of contextual factors on users' performance. For this purpose, we plan to conduct an in-situ experiment in which we collect data from participants in their own real world settings. For data collection, we will use AWARE Framework[2]. AWARE Framework is an accessibility service that runs in the background and collect sensor and interaction data from user device. Moreover, it handles study registration and data synchronization with an external server by applying heuristics on when to syncronize data (based on Wi-Fi connection and battery level). It supports many sensors such as accelerometer, barometer, light, GPS data, screen status as well as active applications or application installations. We are planning to consider GPS and light sensors for location (indoor/outdoor); barometer, gravity, light, magnetometer and temperature sensors for

---

[2]http://www.awareframework.com/.

environmental attributes; accelerometer and gyroscope sensors for mobility and movement; screen and application sensors for temporal context, interruptions and task model; communication and application sensors for interpersonal interaction. It also provides a mechanism to send questions to participants to collect data for specific purposes. While collecting sensor and interaction data from participants, we plan to predict if participants make errors or have problems with the device under specific contextual factors. First, we begin with heuristic rules for error prediction. Then, we will build a mechanism that learns user behaviour under certain contextual settings. When we detect an error, we plan to ask user about the error. At this point, user may confirm the error or refuse it. In both cases, it will be a feedback for our error prediction mechanism. One drawback of this approach is that, there is not a specific task model; as a result, user's intentions will not be known. In order to resolve this issue, we have developed a plugin on AWARE Framework which collects the layout of the user interface whenever it changes. This will help us identify semantic roles of user interface elements that are currently interacted with and cause specific events such as page navigation.

Data set that will be collected during the experiments will be highly sensitive and may arise ethical, privacy and security concerns [13]. Therefore, studies of this kind need to seek for ethical approvals. First, all participants will enroll in our study voluntarily and be free to leave any time they feel uncomfortable with the experiments. They will be informed about the data collected during informed-consent process. Moreover, AWARE Framework provides a user interface to enable/disable data collection from specific sensors or completely disable data logging; so that, users can choose what to submit when they have privacy concerns. Finally, data will be transferred using Secure Sockets Layer (SSL) encryption.

*How can we develop an adaptive system based on interaction and contextual data from users interacting with small screen devices?* After we successfully predict interaction problems in small screen devices, we plan to implement a system that runs in the background and analyze user interaction and sensor data. Whenever a possible interaction problem is predicted, it will send a broadcast message to other applications. Application developers who want to adapt their user interfaces to users and context can integrate their applications with our system by implementing broadcast receivers. The adaptation method depends on task model and interaction problem.

*How does the adaptive system developed in this study affect user performance on small devices?* In order to evaluate our adaptive system, we plan to develop an application that listens for broadcast messages from the system. Unlike experimental applications in the literature, this application will be used for real world tasks. For instance, Telegram[3] is a widely used instant messaging application and its source codes are publicly available. It can be extended with adaptive features for evaluation purposes. The extended application in this phase will satisfy the requirements of ability based approach [33], in terms of ability, accountability, adaptation, transparency, performance, context and commodity. For this purpose, the application will monitor user performance and contextual changes so that it can apply adaptations to improve user performance with respect to users' changing abilities or situational factors. It will also ask user to accept or reject an adaptation, so that users will be aware of the changes on the system. Moreover, these accept or reject responses will be used as a feedback in the performance monitoring system. We plan to conduct an additional in-situ experiment with a different set of participants from different locations. We are going to evaluate the system with respect to feedback from participants and reviews at the end of the experiments.

---

[3] https://telegram.org/

## SUMMARY AND FUTURE DIRECTIONS

The aim of this study is to investigate the effects of contextual factors on user performance in the wild and develop a system which learns and predicts users' abilities and performance changes under different contextual factors to adapt itself to improve user performance. Our systematic review has shown that, although the effect of context on small screen device users' performance has been widely studied, majority of these studies are based on experimental settings and do not reflect real world cases. In this study, we aim to conduct in-situ experiments with real world tasks to better understand how users interact with their small screen device under different contextual factors. One implication of this study is that, it will produce a rich dataset which includes various interaction patterns and associated sensor data. Moreover, the adaptive system that will be developed as part of this study aims to improve user experience by resolving interaction problems due to contextual factors.

## REFERENCES

[1] Ahmed Sabbir Arif, Benedikt Iltisberger, and Wolfgang Stuerzlinger. 2011. Extending Mobile User Ambient Awareness for Nomadic Text Entry. In *Proceedings of the 23rd Australian Computer-Human Interaction Conference (OzCHI '11)*. ACM, New York, NY, USA, 21–30. https://doi.org/10.1145/2071536.2071539

[2] Leon Barnard, Ji Soo Yi, Julie A. Jacko, and Andrew Sears. 2005. An Empirical Comparison of Use-in-motion Evaluation Scenarios for Mobile Computing Devices. *Int. J. Hum.-Comput. Stud.* 62, 4 (April 2005), 487–520. https://doi.org/10.1016/j.ijhcs.2004.12.002

[3] Matthias Böhmer, Brent Hecht, Johannes Schöning, Antonio Krüger, and Gernot Bauer. 2011. Falling Asleep with Angry Birds, Facebook and Kindle: A Large Scale Study on Mobile Application Usage. In *Proceedings of the 13th International Conference on Human Computer Interaction with Mobile Devices and Services (MobileHCI '11)*. ACM, New York, NY, USA, 47–56. https://doi.org/10.1145/2037373.2037383

[4] Andrew Bragdon, Eugene Nelson, Yang Li, and Ken Hinckley. 2011. Experimental Analysis of Touch-screen Gesture Designs in Mobile Environments. In *Proceedings of the SIGCHI Conference on Human Factors in Computing Systems (CHI '11)*. ACM, New York, NY, USA, 403–412. https://doi.org/10.1145/1978942.1979000

[5] Stephen Brewster. 2002. Overcoming the Lack of Screen Space on Mobile Computers. *Personal Ubiquitous Comput.* 6, 3 (Jan. 2002), 188–205. https://doi.org/10.1007/s007790200019

[6] James Clawson, Thad Starner, Daniel Kohlsdorf, David P. Quigley, and Scott Gilliland. 2014. Texting While Walking: An Evaluation of Mini-qwerty Text Input While On-the-go. In *Proceedings of the 16th International Conference on Human-computer Interaction with Mobile Devices & Services (MobileHCI '14)*. ACM, New York, NY, USA, 339–348. https:

//doi.org/10.1145/2628363.2628408
[7] Jessica Conradi. 2017. Influence of Letter Size on Word Reading Performance During Walking. In *Proceedings of the 19th International Conference on Human-Computer Interaction with Mobile Devices and Services (MobileHCI '17)*. ACM, New York, NY, USA, Article 16, 9 pages. https://doi.org/10.1145/3098279.3098554
[8] Anind K. Dey, Gregory D. Abowd, and Daniel Salber. 2001. A Conceptual Framework and a Toolkit for Supporting the Rapid Prototyping of Context-aware Applications. *Hum.-Comput. Interact.* 16, 2 (Dec. 2001), 97–166. https://doi.org/10.1207/S15327051HCI16234_02
[9] David Dobbelstein, Gabriel Haas, and Enrico Rukzio. 2017. The Effects of Mobility, Encumbrance, and (Non-)Dominant Hand on Interaction with Smartwatches. In *Proceedings of the 2017 ACM International Symposium on Wearable Computers (ISWC '17)*. ACM, New York, NY, USA, 90–93. https://doi.org/10.1145/3123021.3123033
[10] Shimin Feng, Graham Wilson, Alex Ng, and Stephen Brewster. 2015. Investigating Pressure-based Interactions with Mobile Phones While Walking and Encumbered. In *Proceedings of the 17th International Conference on Human-Computer Interaction with Mobile Devices and Services Adjunct (MobileHCI '15)*. ACM, New York, NY, USA, 854–861. https://doi.org/10.1145/2786567.2793711
[11] Stephen Fitchett and Andy Cockburn. 2009. Evaluating Reading and Analysis Tasks on Mobile Devices: A Case Study of Tilt and Flick Scrolling. In *Proceedings of the 21st Annual Conference of the Australian Computer-Human Interaction Special Interest Group: Design: Open 24/7 (OZCHI '09)*. ACM, New York, NY, USA, 225–232. https://doi.org/10.1145/1738826.1738863
[12] Daniel Fitton, I. Scott MacKenzie, Janet C. Read, and Matthew Horton. 2013. Exploring Tilt-based Text Input for Mobile Devices with Teenagers. In *Proceedings of the 27th International BCS Human Computer Interaction Conference (BCS-HCI '13)*. British Computer Society, Swinton, UK, UK, Article 25, 6 pages. http://dl.acm.org/citation.cfm?id=2578048.2578082
[13] Gabriella M. Harari, Nicholas D. Lane, Rui Wang, Benjamin S. Crosier, Andrew T. Campbell, and Samuel D. Gosling. 2016. Using Smartphones to Collect Behavioral Data in Psychological Science: Opportunities, Practical Considerations, and Challenges. *Perspectives on Psychological Science* 11, 6 (2016), 838–854. https://doi.org/10.1177/1745691616650285 arXiv:https://doi.org/10.1177/1745691616650285 PMID: 27899727.
[14] Simon Harper, Tianyi Chen, and Yeliz Yesilada. 2011. How Do People Use Their Mobile Phones?: A Field Study of Small Device Users. *Int. J. Mob. Hum. Comput. Interact.* 3, 1 (Jan. 2011), 37–54. https://doi.org/10.4018/jmhci.2011010103
[15] Simon Harper, Yeliz Yesilada, and Tianyi Chen. 2011. Mobile device impairment ... similar problems, similar solutions? *Behaviour & Information Technology* 30, 5 (2011), 673–690. https://doi.org/10.1080/01449291003801943 arXiv:https://doi.org/10.1080/01449291003801943
[16] Morgan Harvey and Matthew Pointon. 2017. Perceptions of the Effect of Fragmented Attention on Mobile Web Search Tasks. In *Proceedings of the 2017 Conference on Conference Human Information Interaction and Retrieval (CHIIR '17)*. ACM, New York, NY, USA, 293–296. https://doi.org/10.1145/3020165.3022136
[17] Eve Hoggan, Andrew Crossan, Stephen A. Brewster, and Topi Kaaresoja. 2009. Audio or Tactile Feedback: Which Modality when?. In *Proceedings of the SIGCHI Conference on Human Factors in Computing Systems (CHI '09)*. ACM, New York, NY, USA, 2253–2256. https://doi.org/10.1145/1518701.1519045
[18] Ira E. Hyman, S. Matthew Boss, Breanne M. Wise, Kira E. McKenzie, and Jenna M. Caggiano. 2010. Did you see the unicycling clown? Inattentional blindness while walking and talking on a cell phone. *Applied Cognitive Psychology* 24, 5 (2010), 597–607. https://doi.org/10.1002/acp.1638
[19] Satu Jumisko-Pyykkö and Teija Vainio. 2010. Framing the Context of Use for Mobile HCI. *Int. J. Mob. Hum. Comput. Interact.* 2, 4 (Oct. 2010), 1–28. https://doi.org/10.4018/jmhci.2010100101
[20] Eric M. Lamberg and Lisa M. Muratori. 2012. Cell phones change the way we walk. *Gait & Posture* 35, 4 (2012), 688 – 690. https://doi.org/10.1016/j.gaitpost.2011.12.005

[21] Bonnie MacKay, David Dearman, Kori Inkpen, and Carolyn Watters. 2005. Walk 'N Scroll: A Comparison of Software-based Navigation Techniques for Different Levels of Mobility. In *Proceedings of the 7th International Conference on Human Computer Interaction with Mobile Devices &Amp; Services (MobileHCI '05)*. ACM, New York, NY, USA, 183–190. https://doi.org/10.1145/1085777.1085808

[22] Alexander Ng, John Williamson, and Stephen Brewster. 2015. The Effects of Encumbrance and Mobility on Touch-Based Gesture Interactions for Mobile Phones. In *Proceedings of the 17th International Conference on Human-Computer Interaction with Mobile Devices and Services (MobileHCI '15)*. ACM, New York, NY, USA, 536–546. https://doi.org/10.1145/2785830.2785853

[23] Antti Oulasvirta, Sakari Tamminen, Virpi Roto, and Jaana Kuorelahti. 2005. Interaction in 4-second Bursts: The Fragmented Nature of Attentional Resources in Mobile HCI. In *Proceedings of the SIGCHI Conference on Human Factors in Computing Systems (CHI '05)*. ACM, New York, NY, USA, 919–928. https://doi.org/10.1145/1054972.1055101

[24] Keith B. Perry and Juan Pablo Hourcade. 2008. Evaluating One Handed Thumb Tapping on Mobile Touchscreen Devices. In *Proceedings of Graphics Interface 2008 (GI '08)*. Canadian Information Processing Society, Toronto, Ont., Canada, Canada, 57–64. http://dl.acm.org/citation.cfm?id=1375714.1375725

[25] Markus Pierer. 2016. *Mobile Device types*. Springer Fachmedien Wiesbaden, Wiesbaden, 31–36. https://doi.org/10.1007/978-3-658-15046-4_4

[26] Zhanna Sarsenbayeva, Jorge Goncalves, Juan García, Simon Klakegg, Sirkka Rissanen, Hannu Rintamäki, Jari Hannu, and Vassilis Kostakos. 2016. Situational Impairments to Mobile Interaction in Cold Environments. In *Proceedings of the 2016 ACM International Joint Conference on Pervasive and Ubiquitous Computing (UbiComp '16)*. ACM, New York, NY, USA, 85–96. https://doi.org/10.1145/2971648.2971734

[27] Andrew Sears and Mark Young. 2003. The Human-computer Interaction Handbook. L. Erlbaum Associates Inc., Hillsdale, NJ, USA, Chapter Physical Disabilities and Computing Technologies: An Analysis of Impairments, 482–503. http://dl.acm.org/citation.cfm?id=772072.772105

[28] Lin M. Jacko J. Sears, A. and Y Xiao. 2003. When Computers Fade . . . Pervasive Computing and Situationally-Induced Impairments and Disabilities. In *Proceedings of HCII 2003*. 1298–1302.

[29] Despina Stavrinos, Katherine W. Byington, and David C. Schwebel. 2011. Distracted walking: Cell phones increase injury risk for college pedestrians. *Journal of Safety Research* 42, 2 (2011), 101 – 107. https://doi.org/10.1016/j.jsr.2011.01.004

[30] Kristin Vadas, Nirmal Patel, Kent Lyons, Thad Starner, and Julie Jacko. 2006. Reading On-the-go: A Comparison of Audio and Hand-held Displays. In *Proceedings of the 8th Conference on Human-computer Interaction with Mobile Devices and Services (MobileHCI '06)*. ACM, New York, NY, USA, 219–226. https://doi.org/10.1145/1152215.1152262

[31] Jyh-Da Wei, Hsu-Fu Hsiao, and Pei-Yu Jiang. 2016. A System Modeling Based Anti-Shake Technique for Mobile Display. In *Proceedings of the 2016 CHI Conference Extended Abstracts on Human Factors in Computing Systems (CHI EA '16)*. ACM, New York, NY, USA, 3241–3246. https://doi.org/10.1145/2851581.2892514

[32] Graham Wilson, Stephen A. Brewster, Martin Halvey, Andrew Crossan, and Craig Stewart. 2011. The Effects of Walking, Feedback and Control Method on Pressure-based Interaction. In *Proceedings of the 13th International Conference on Human Computer Interaction with Mobile Devices and Services (MobileHCI '11)*. ACM, New York, NY, USA, 147–156. https://doi.org/10.1145/2037373.2037397

[33] Jacob O. Wobbrock, Shaun K. Kane, Krzysztof Z. Gajos, Susumu Harada, and Jon Froehlich. 2011. Ability-Based Design: Concept, Principles and Examples. *ACM Trans. Access. Comput.* 3, 3, Article 9 (April 2011), 27 pages. https://doi.org/10.1145/1952383.1952384

[34] Yeliz Yesilada, Giorgio Brajnik, and Simon Harper. 2011. Barriers Common to Mobile and Disabled Web Users. *Interacting With Computers* 23, 5 (2011), 525–542. https://doi.org/10.1016/j.intcom.2011.05.005